\documentclass[%
reprint,
superscriptaddress,
 amsmath,amssymb,
 aps,
 prb,
]{revtex4-2}

\usepackage{graphicx}
\usepackage{dcolumn}
\usepackage{bm}

\usepackage{color}
\usepackage{ulem}

\begin{document}


\title{Evolution of Band Structure in a Kagome Superconductor $\textrm{Cs(V}_{1-x}\textrm{Cr}_{x})_{3}\textrm{Sb}_{5}$: Toward Universal Understanding of CDW and Superconducting Phase Diagrams}

\author{Shuto Suzuki}
\thanks{These authors contributed equally to this work.}
\affiliation{Department of Physics, Graduate School of Science, Tohoku University, Sendai 980-8578, Japan}

\author{Takemi Kato}
\thanks{These authors contributed equally to this work.}
\affiliation{Advanced Institute for Materials Research (WPI-AIMR), Tohoku University, Sendai 980-8577, Japan}

\author{Yongkai Li}
\thanks{These authors contributed equally to this work.}
\affiliation{Centre for Quantum Physics, Key Laboratory of Advanced Optoelectronic Quantum Architecture and Measurement (MOE), School of Physics, Beijing Institute of Technology, Beijing 100081, P. R. China}
\affiliation{Beijing Key Lab of Nanophotonics and Ultrafine Optoelectronic Systems, Beijing Institute of Technology, Beijing 100081, P. R. China}
\affiliation{Material Science Center, Yangtze Delta Region Academy of Beijing Institute of Technology, Jiaxing, 314011, P. R. China}

\author{Kosuke Nakayama}
\thanks{Corresponding authors:\\
k.nakayama@arpes.phys.tohoku.ac.jp\\
zhiweiwang@bit.edu.cn\\
t-sato@arpes.phys.tohoku.ac.jp}
\affiliation{Department of Physics, Graduate School of Science, Tohoku University, Sendai 980-8578, Japan}

\author{Zhiwei Wang}
\thanks{Corresponding authors:\\
k.nakayama@arpes.phys.tohoku.ac.jp\\
zhiweiwang@bit.edu.cn\\
t-sato@arpes.phys.tohoku.ac.jp}
\affiliation{Centre for Quantum Physics, Key Laboratory of Advanced Optoelectronic Quantum Architecture and Measurement (MOE), School of Physics, Beijing Institute of Technology, Beijing 100081, P. R. China}
\affiliation{Beijing Key Lab of Nanophotonics and Ultrafine Optoelectronic Systems, Beijing Institute of Technology, Beijing 100081, P. R. China}
\affiliation{Material Science Center, Yangtze Delta Region Academy of Beijing Institute of Technology, Jiaxing, 314011, P. R. China}
\affiliation{Beijing Institute of Technology, Zhuhai, 519000, P. R. China}

\author{Seigo Souma}
\affiliation{Advanced Institute for Materials Research (WPI-AIMR), Tohoku University, Sendai 980-8577, Japan}
\affiliation{Center for Science and Innovation in Spintronics (CSIS), Tohoku University, Sendai 980-8577, Japan}

\author{Kenichi Ozawa}
\affiliation{Institute of Materials Structure Science, High Energy Accelerator Research Organization (KEK), Tsukuba, Ibaraki 305-0801, Japan}

\author{Miho Kitamura}
\affiliation{Institute of Materials Structure Science, High Energy Accelerator Research Organization (KEK), Tsukuba, Ibaraki 305-0801, Japan}
\affiliation{National Institutes for Quantum Science and Technology (QST), Sendai 980-8579, Japan}

\author{Koji Horiba}
\affiliation{National Institutes for Quantum Science and Technology (QST), Sendai 980-8579, Japan}

\author{Hiroshi Kumigashira}
\affiliation{Institute of Multidisciplinary Research for Advanced Materials (IMRAM), Tohoku University, Sendai 980-8577, Japan}

\author{Takashi Takahashi}
\affiliation{Department of Physics, Graduate School of Science, Tohoku University, Sendai 980-8578, Japan}

\author{Yugui Yao}
\affiliation{Centre for Quantum Physics, Key Laboratory of Advanced Optoelectronic Quantum Architecture and Measurement (MOE), School of Physics, Beijing Institute of Technology, Beijing 100081, P. R. China}
\affiliation{Beijing Key Lab of Nanophotonics and Ultrafine Optoelectronic Systems, Beijing Institute of Technology, Beijing 100081, P. R. China}

\author{Takafumi Sato}
\thanks{Corresponding authors:\\
k.nakayama@arpes.phys.tohoku.ac.jp\\
zhiweiwang@bit.edu.cn\\
t-sato@arpes.phys.tohoku.ac.jp}
\affiliation{Department of Physics, Graduate School of Science, Tohoku University, Sendai 980-8578, Japan}
\affiliation{Advanced Institute for Materials Research (WPI-AIMR), Tohoku University, Sendai 980-8577, Japan}
\affiliation{Center for Science and Innovation in Spintronics (CSIS), Tohoku University, Sendai 980-8577, Japan}
\affiliation{International Center for Synchrotron Radiation Innovation Smart (SRIS), Tohoku University, Sendai 980-8577, Japan}
\affiliation{Mathematical Science Center for Co-creative Society (MathCCS), Tohoku University, Sendai 980-8578, Japan}

\date{\today}

\begin{abstract}
Kagome superconductors $A\textrm{V}_{3}\textrm{Sb}_{5}$ ($A =$ K, Rb, Cs) exhibit a characteristic superconducting and charge-density wave (CDW) phase diagram upon carrier doping and chemical substitution. However, the key electronic states responsible for such a phase diagram have yet to be clarified.
Here we report a systematic micro-focused angle-resolved photoemission spectroscopy (ARPES) study of $\textrm{Cs(V}_{1-x}\textrm{Cr}_{x})_{3}\textrm{Sb}_{5}$ as a function of Cr content $x$, where Cr substitution causes monotonic reduction of superconducting and CDW transition temperatures.
We found that the V-derived bands forming saddle points at the M point and Dirac nodes along high-symmetry cuts show an energy shift due to electron doping by Cr substitution, whereas the Sb-derived electron band at the $\Gamma$ point remains almost unchanged, signifying an orbital-selective band shift.
We also found that band doubling associated with the emergence of three-dimensional CDW identified at $x = 0$ vanishes at $x \ge 0.25$, in line with the disappearance of CDW.
A comparison of band diagrams among Ti-, Nb-, and Cr-substituted $\textrm{Cs}\textrm{V}_{3}\textrm{Sb}_{5}$ suggests the importance to simultaneously take into account the two saddle points at the M point and their proximity to the Fermi energy, to understand the complex phase diagram against carrier doping and chemical pressure.
\end{abstract}

\maketitle

Two-dimensional (2D) kagome lattice provides a fertile ground for studying exotic quantum phenomena originating from correlated and topological band structures, featuring a Dirac-cone band, a flat band, and a saddle-point (SP) band \cite{WangNatRevPhys2023}.
Among these bands, the SP band forms a van Hove singularity, leading to a divergence in the density of states (DOS).
This enhances many-body interactions and Fermi-surface (FS) instability, inducing various ordered phases, such as unconventional density waves, superconductivity, and magnetism \cite{YuPRB2012, KieselPRB2012, WangPRB2013, KieselPRL2013}.
However, achieving such phases requires a tuning of the SP band close to the Fermi energy ($E_{\textrm{F}}$), which has been challenging in kagome lattices.

Recently, $A\textrm{V}_{3}\textrm{Sb}_{5}$ ($A =$ K, Rb, Cs) [Fig. 1(a)] has been identified as the first kagome-lattice family with electron filling near the SP band \cite{OrtizPRM2019}.
Subsequent experiments have confirmed the coexistence of multiple SP bands near $E_{\textrm{F}}$.
$A\textrm{V}_{3}\textrm{Sb}_{5}$ exhibits various competing or intertwined orders including CDW, superconductivity, pair density wave, nematicity, and time-reversal-symmetry breaking \cite{OrtizPRL2020, OrtizPRM2021, YinCPL2021, ChenNature2021, ZhaoNature2021, JiangNM2021, ShumiyaPRB2021, WangPRB2021, NieNature2022, MielkeNature2022, XuNP2022, JiangNanoLett2023}.
The relationship of these ordered states to the unique band structure in $A\textrm{V}_{3}\textrm{Sb}_{5}$ is a matter of intensive debate.

A key strategy to address this issue is to apply perturbations (e.g., pressure, elemental substitution, and carrier doping \cite{DuPRB2021, ChenPRL2021, ZhangPRB2021, DuCPB2021, SongPRL2021, YangSB2022, LiuPRM2023, LiPRB2022, KatoPRL2022, ZhouPRB2023, OeyPRM2022, OeyPRM2022_2, DingPRB2022, PengarXiv2024, ZhongNCOM2023, ZhongNature2023, LuoNCOM2023, HuainpjQM2024, LiuPRB2022, LiuSciRep2024, QianPRB2021, ZhuPRR2024, NakayamaPRX2022, PengPRB2024}) that presumably tune the band structure and the ordered states, to clarify their relationship.
Recent experiments on $\textrm{Cs}\textrm{V}_{3}\textrm{Sb}_{5}$ revealed intriguing phase diagrams resulting from such perturbations.
For instance, isovalent substitution of V with Nb or Ta was shown to decrease the CDW transition temperature ($T_{\textrm{CDW}}$) while increasing the superconducting transition temperature ($T_{\textrm{c}}$) \cite{LiPRB2022, LiuPRB2022, LiuSciRep2024}.
In contrast, substituting V with Ti to dope hole carriers initially decreases $T_{\textrm{c}}$ as $T_{\textrm{CDW}}$ is lowered, and then begins to increase $T_{\textrm{c}}$, forming a two-dome-like superconducting state \cite{YangSB2022, LiuPRM2023}.
More recently, $\textrm{Cs}\textrm{V}_{3}\textrm{Sb}_{5}$ samples with Cr substitution, introducing electron carriers, was reported to show reduction of both CDW and superconductivity [Fig. 1(b)] \cite{DingPRB2022, PengarXiv2024}.
Given these remarkable differences in the phase diagram, an experimental study on the underlying electronic states would provide valuable information into the key electronic states responsible for CDW and superconductivity.
However, such studies are scarce, especially regarding the Cr-substituted case.
This limits comprehensive understanding, such as whether CDW is commonly suppressed when SPs are shifted away from $E_{\textrm{F}}$.

In this article, we report a systematic ARPES study of Cr-substituted $\textrm{Cs}\textrm{V}_{3}\textrm{Sb}_{5}$, $\textrm{Cs(V}_{1-x}\textrm{Cr}_{x})_{3}\textrm{Sb}_{5}$ ($x = 0, 0.25$, and 0.45).
We found an orbital-selective band shift upon Cr substitution, characterized by a downward shift of V-derived SP and Dirac-cone bands while pinning the energy position of the Sb-derived electron bands at the $\Gamma$ point.
We also found that the appearance of CDW gap and band doubling is well correlated with a reduction of $T_{\textrm{CDW}}$.
We discuss implications of the present results in relation to the mechanism of CDW and superconductivity, and further discuss the role of chemical substitution to the band diagram through a comparison with Ti- and Nb-substituted $\textrm{Cs}\textrm{V}_{3}\textrm{Sb}_{5}$.

High-quality single crystals of $\textrm{Cs(V}_{1-x}\textrm{Cr}_{x})_{3}\textrm{Sb}_{5}$ were synthesized with the self-flux method \cite{PengarXiv2024}.
ARPES measurements were performed using a Scienta-Omicron DA30 spectrometer at BL-28A in Photon Factory, KEK, with a micro-focused beam spot of $10 \times12$ $\mu$m$^2$ \cite{KitamuraRSI2022}.
Circularly polarized  photons with photon energy of $h\nu = 106$ eV were used to probe the $ k_z = 0$ ($\Gamma$KM) plane \cite{NakayamaPRB2021}.
Samples were cleaved \textit{in}-\textit{situ} at low temperatures and the surface termination (Sb vs Cs) was determined by scanning surface area and monitoring core-level intensity as detailed below.
Energy resolution was set to be 35 meV.

\begin{figure}[htbp]
\includegraphics[width=86mm]{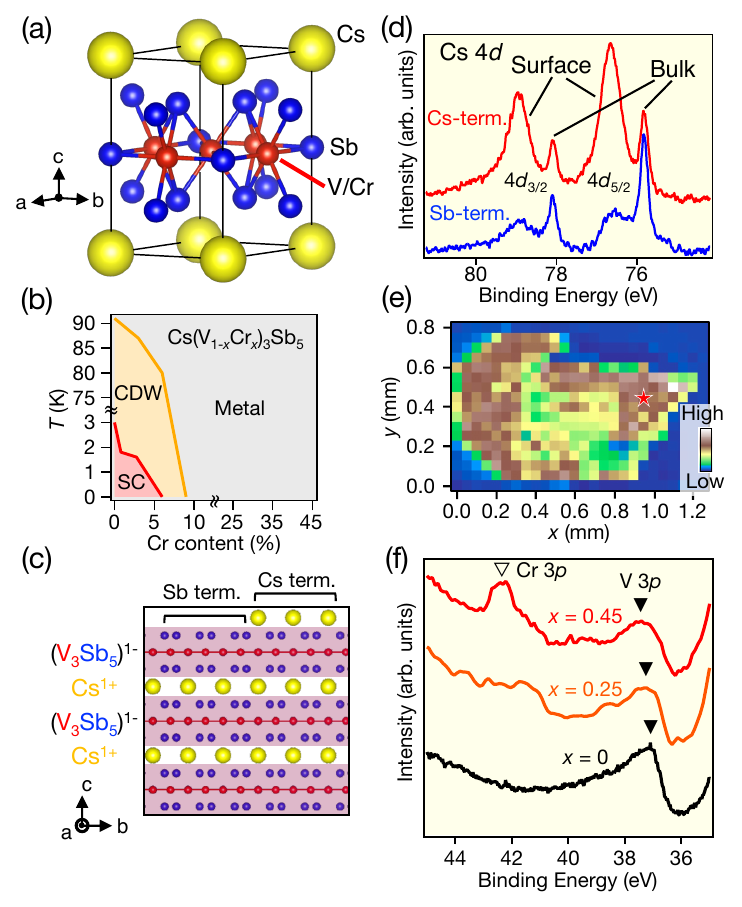}
 \caption{\label{Fig1}
(a) Crystal structure of $\textrm{Cs(V}_{1-x}\textrm{Cr}_{x})_{3}\textrm{Sb}_{5}$.
The graphic was created by VESTA \cite{MommaJApplCryst2011}.
(b) CDW and superconducting phase diagram of $\textrm{Cs(V}_{1-x}\textrm{Cr}_{x})_{3}\textrm{Sb}_{5}$ as a function of $x$ and temperature \cite{PengarXiv2024, DingPRB2022}.
(c) Side view of crystal with cleaved surface terminated by either Sb or Cs atoms.
(d) EDCs in the Cs $4d$ core-level region for Cs- and Sb-terminated surfaces at $x = 0.45$.
(e) Spatial image of the core-level intensity on a cleaved surface for the Cs $4d$ surface peak.
Red star indicates the location where ARPES measurements shown in Figs. 2 and 3 were carried out.
(f) EDCs in the V/Cr $3p$ core-level region.}
 \end{figure}

Previous scanning tunneling microscopy (STM) and spatially-resolved ARPES studies \cite{ChenNature2021, ZhaoNature2021, JiangNM2021, ShumiyaPRB2021, WangPRB2021, LiangPRX2021, YuNanoLett2022, KatoPRB2022, LuoPRB2022, KatoPRB2023, JiangNanoLett2023} suggested that the cleaved surface of $A\textrm{V}_{3}\textrm{Sb}_{5}$ contains Sb and $A$ terminations [Fig. 1(c)].
The CDW properties are markedly different between the two terminations; the $A$ termination reflects bulk-originated CDW whereas the Sb termination shows no indication of a CDW gap.
It is thus essential to perform surface-termination-selective ARPES measurements also for $\textrm{Cs(V}_{1-x}\textrm{Cr}_{x})_{3}\textrm{Sb}_{5}$.
As shown in Fig. 1(d), we distinguished Sb- and Cs-terminated surfaces in the Cr-substituted samples based on the knowledge that the spectral weight of the surface-derived Cs core levels is enhanced for the Cs-terminated surface (red curve) relative to the Sb-terminated one (blue curve).
We carefully selected the Cs-terminated surface by scanning the whole area of cleaved surface [see representative mapping image in Fig. 1(e)].
To elucidate the systematic evolution upon Cr substitution, it is also essential to clarify whether the Cr atoms indeed exist at the cleaved surface by ARPES.
As shown in Fig. 1(f), after Cr substitution, the Cr $3p$ core-level peak appears at the binding energy ($E_{\textrm{B}}$) of $\sim$42 eV, and its intensity for $x = 0.45$ becomes nearly comparable to that of the V $3p$ peak, consistent with the fact that almost a half of V atoms are replaced with Cr atoms.

\begin{figure*}[htbp]
\includegraphics[width=165mm]{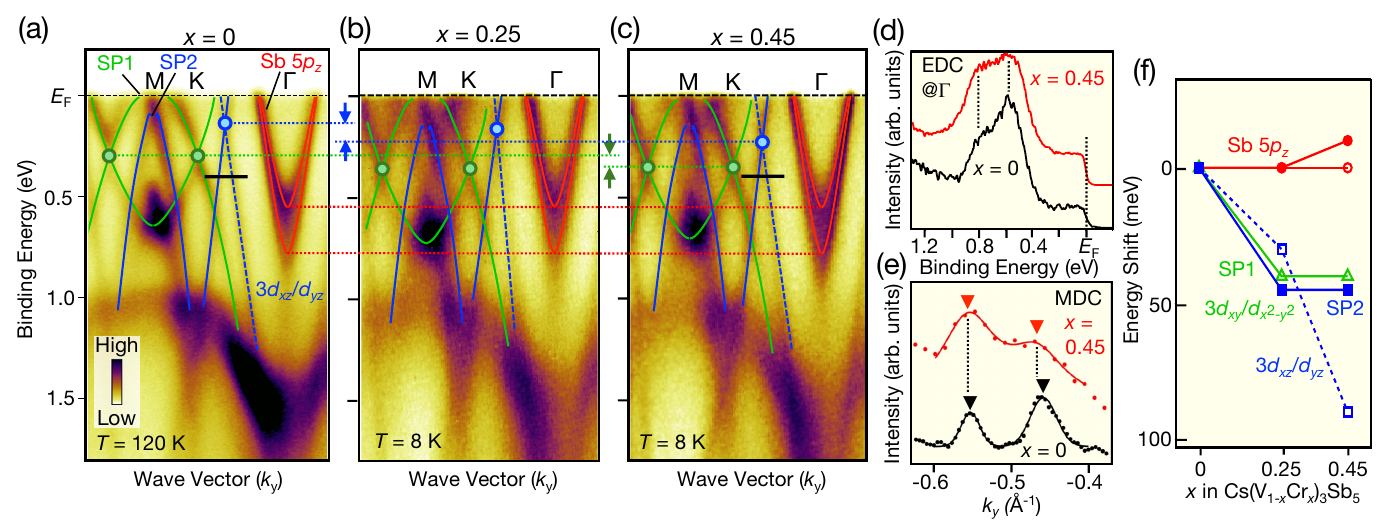}
 \caption{\label{Fig2}
(a)--(c) ARPES intensity of $\textrm{Cs(V}_{1-x}\textrm{Cr}_{x})_{3}\textrm{Sb}_{5}$ along the $\Gamma$KM cut for $x = 0$, 0.25, and 0.45, respectively, measured in the normal state ($T = 120$ K for $x = 0$, and $T = 8$ K for $x = 0.25$ and 0.45) for the Cs-terminated surface.
Solid curves are a guide for the eyes to trace the band dispersion.
Horizontal dashed lines and arrows highlight the energy difference of each band among different $x$.
(d) Comparison of EDC at the $\Gamma$ point between $x = 0$ and 0.45.
(e) Comparison of MDC between $x = 0$ and 0.45 along cuts shown by black lines in (a) and (c) (red and black dots, respectively), together with the results of numerical fittings (solid curves).
(f) Energy shift of Sb $5p_{z}$, SP1, SP2, and V/Cr $3d_{xz}/d_{yz}$ bands upon Cr substitution relative to the band position for $x$ = 0, estimated by analyzing the peak position of EDCs.}
\end{figure*}

Next we present the evolution of the band structure upon Cr substitution.
Figures 2(a)--2(c) show a comparison of the band dispersion above $T_{\textrm{CDW}}$ among pristine ($x = 0$) and Cr-substituted ($x = 0.25$ and 0.45) $\textrm{Cs}\textrm{V}_{3}\textrm{Sb}_{5}$, measured along the $\Gamma$KM cut of bulk hexagonal Brillouin zone.
At $x = 0$ [Fig. 2(a)], one can immediately recognize several dispersive bands.
For example, (i) highly dispersive electron bands at the $\Gamma$ point (red curves), (ii) linearly dispersive bands that intersect each other midway between the $\Gamma$ and K points (blue solid and dashed lines), (iii) a holelike band forming a SP (called SP1) at the M point slightly above $E_{\textrm{F}}$ which intersects the M-centered electronlike band at the K point (green curves), and (iv) another holelike band topped at the M point slightly below $E_{\textrm{F}}$ (blue curve) which forms another SP (called SP2).
These bands are attributed to the (i) Sb $5p_{z}$ and (ii-iv) V $3d$ kagome-lattice bands (namely, green, solid blue, and dashed blue curves represent dominant contributions from V $3d_{xy}/d_{x^{2}-y^{2}}$, V $3d_{xz}/d_{yz}$, and another V $3d_{xz}/d_{yz}$ orbitals, respectively) \cite{JiangNM2021, TanPRL2021, WangPRB2022, FuPRB2021, ZhaoNature2021, NakayamaPRB2021, KatoCOMMAT2022, HuNCOM2022, KangNP2022}.
All these bands are commonly observed in the Cr-substituted samples [Figs. 2(b) and 2(c)], indicating that the 45\% Cr doping does not qualitatively alter the overall valence-band structure.
However, a careful look at Figs. 2(a)--2(c) reveals a quantitative difference in the band energies upon Cr substitution.
For example, the crossing point associated with the V $3d_{xz}/d_{yz}$ band (blue circles) initially located at $E_{\textrm{B}}$ = 0.15 eV at $x = 0$ moves downward and reaches 0.25 eV at $x = 0.45$.
This trend is also seen for the Dirac point formed by the SP1 band and an electron band (green circles) and also for the $\Lambda$-shaped SP2 band (blue curve) although the total band shift is smaller than that of the V $3d_{xz}/d_{yz}$ band (blue solid and dashed lines).
Interestingly, despite such a downward shift of the V-derived bands, the bottom of the Sb-derived band at the $\Gamma$ point is stationary to the $x$ variation (red curves and red dashed lines).
This can be seen from the EDCs at the $\Gamma$ point for $x = 0$ and 0.45 in Fig. 2(d) which signifies no apparent difference in the peak position.
On the other hand, a direct comparison of the MDCs along a $\textbf{k}$ cut crossing the lower Dirac-cone-like band [black solid lines in Figs. 2(a) and 2(c)] in Fig. 2(e) reveals a finite difference in the peak position between $x = 0$ and 0.45.
The estimated energy shift of each band  [Fig. 2(f)] demonstrates a strong orbital dependence of the total band shift upon $x$ variation; it is nearly zero for the Sb $5p_{z}$ band, $\sim$50 meV for the SP1 and SP2 bands, and $\sim$100 meV for the V $3d_{xz}/d_{yz}$ bands.
The observed downward shift of the V-derived bands upon increasing $x$ would naturally be explained in terms of the electron doping associated with the replacement of group-V element (V) with group-VI element (Cr), which affects V orbitals more than Sb orbitals.

\begin{figure}[htbp]
\includegraphics[width=86mm]{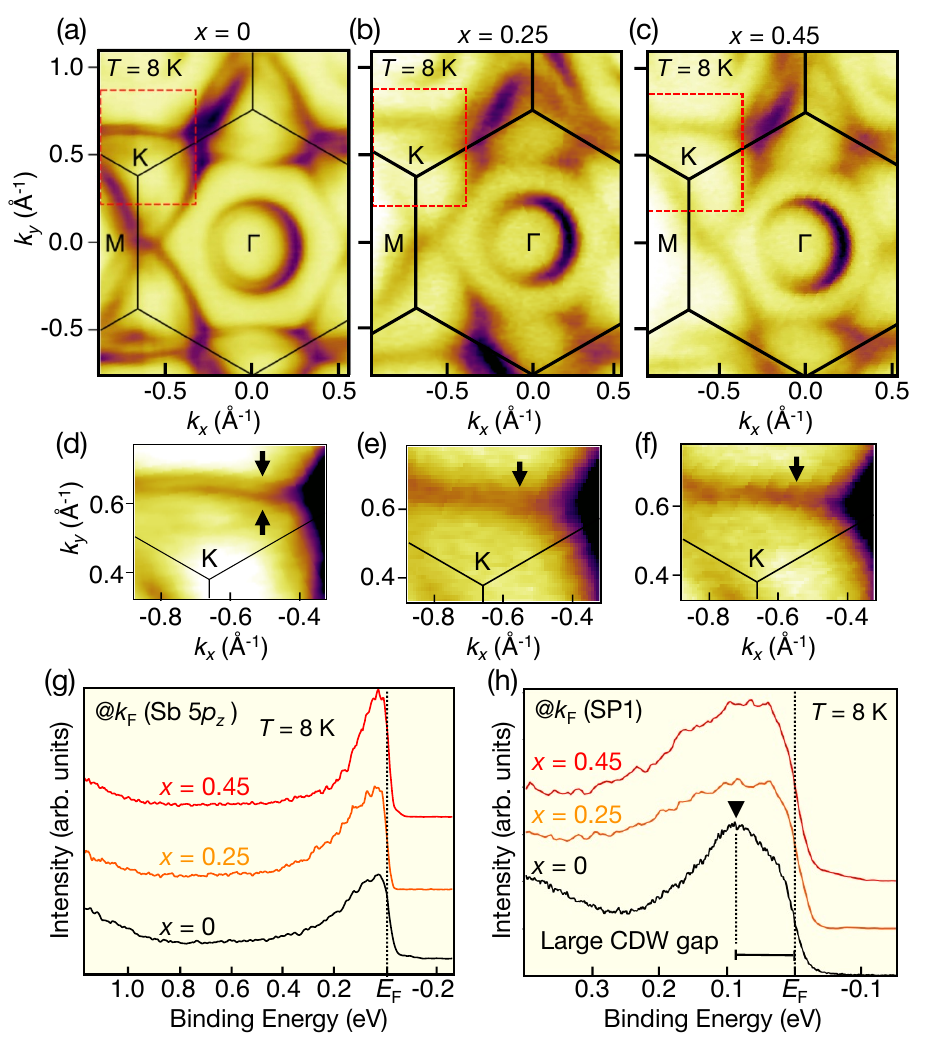}
 \caption{\label{Fig3}
(a)--(c) FS mapping for $x = 0$, 0.25, and 0.45, respectively, measured at $T = 8$ K.
(d)--(f) Enlarged view in the $k$ region enclosed by a dashed rectangle in (a)--(c), respectively.
(g) Comparison of EDCs at $T = 8$ K among $x = 0$, 0.25 and 0.45, obtained at the $\textbf{k}_{\mathrm{F}}$ point of the Sb $5p_{z}$ band.
(h) Same as (g) but of the SP1 band.}
\end{figure}

Now that the overall band evolution upon Cr substitution is established, next we present the influence of CDW to the electronic structure.
Figures 3(a)--3(c) compare the FS mappings at $k_{z} \sim 0$ plane for different $x$, measured at $T =$ 8 K (the CDW state for $x = 0$ and the normal state for $x = 0.25$ and 0.45).
At $x = 0$ [Fig. 3(a)], there are a small circular and a large hexagonal pocket each centered at the $\Gamma$ point, together with a triangular pocket at the K point \cite{LiPRX2021, LouPRL2022, NakayamaPRB2021, LiuPRX2021, ChoPRL2021, LuoNCOM2022, HuNCOM2022, KangNP2022, KatoCOMMAT2022} which are assigned to the Sb $5p_{z}$, V $3d_{xz}/d_{yz}$, and V $3d_{xy}/d_{x^{2}-y^{2}}$ orbitals, respectively (note that a circular pocket splits into two pockets \cite{KatoPRB2022, KatoPRB2023, CaiCOMMAT2024, LuoPRB2022}).
Upon increasing $x$, the hexagonal pocket gradually shrinks whereas the circular pocket keeps the same volume, confirming the orbital-selective band shift.
An expanded view in Fig. 3(d) for $x = 0$ signifies that the hexagonal pocket shows a doubling as indicated by black arrows, likely due to the CDW-induced band folding along the $k_{z}$ direction \cite{KatoPRB2022, KatoPRB2023, KangNM2023, HuPRB2022, LiPRR2022} accompanied by the $2\times2\times2$ out-of-plane unit-cell doubling \cite{LiangPRX2021, LiPRX2021}.
Intriguingly, such a doubling is absent for $x = 0.25$ [Fig. 3(e)] and $x = 0.45$ [Fig. 3(f)] in the non-CDW phase, supporting that the reported band doubling \cite{KatoPRB2022, KatoPRB2023, KangNM2023, HuPRB2022, LiPRR2022} is indeed of CDW origin.

We further examined the influence of CDW to the low-energy excitations by inspecting the EDCs.
Figures 3(g) and 3(h) show EDCs at $T = 8$ K compared among $x = 0$, 0.25 and 0.45, obtained at the Fermi-wave-vector ($\textbf{k}_{\mathrm{F}}$) points of the Sb-$5p_{z}$ band and the SP1 band, respectively.
One can immediately recognize in Fig. 3(g) a sharp peak near $E_{\textrm{F}}$, being cut off by the Fermi-Dirac distribution function.
The leading-edge midpoint is always located at $E_{\textrm{F}}$, indicative of the absence of CDW gap, consistent with previous reports \cite{NakayamaPRB2021, LuoNCOM2022, KangNP2022}.
On the other hand, at the $\textbf{k}_{\mathrm{F}}$ point of the SP1 band [Fig. 3(h)] for $x = 0$, one can recognize a hump feature at $\sim$0.1 eV which is attributed to the large CDW gap \cite{NakayamaPRB2021, KatoCOMMAT2022}.
In sharp contrast, the CDW gap is absent for $x = 0.25$ and 0.45, again consistent with the disappearance of CDW.

\begin{figure}[htbp]
\includegraphics[width=86mm]{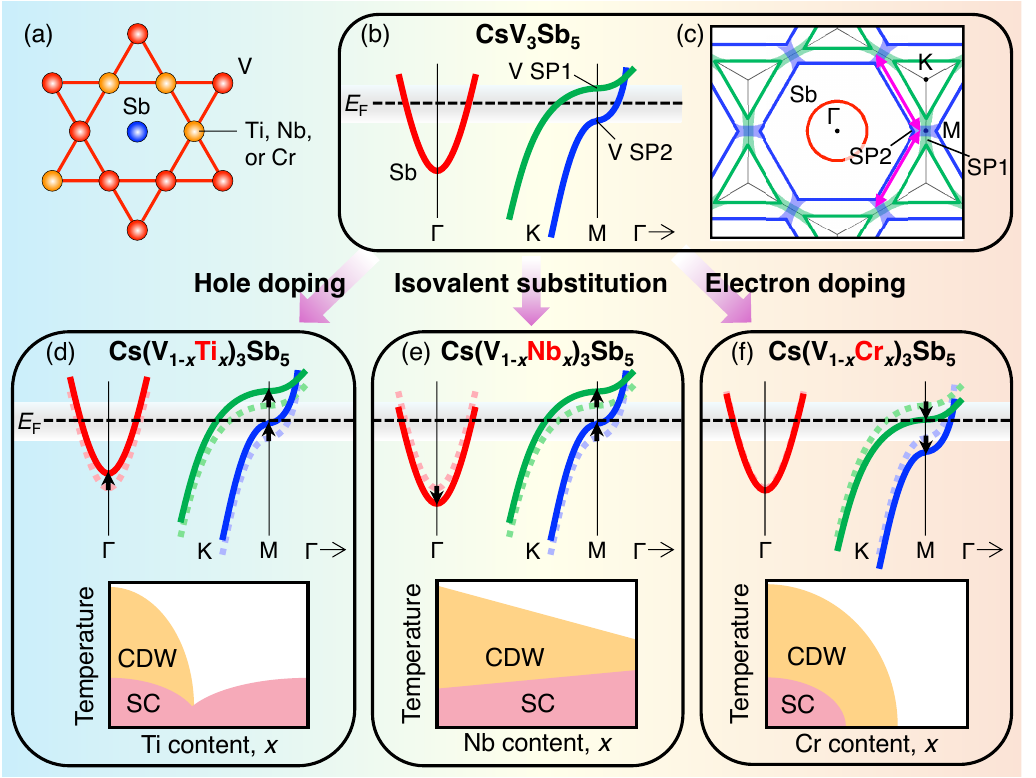}
 \caption{\label{Fig4}
(a) Schematic of the kagome lattice where V atoms are partly substituted with other transition metals (Ti, Nb, or Cr).
(b) Band diagrams for Sb, SP1, and SP2 bands in pristine $\textrm{Cs}\textrm{V}_{3}\textrm{Sb}_{5}$.
(c) Schematic FS of $\textrm{Cs}\textrm{V}_{3}\textrm{Sb}_{5}$.
(d)--(f) Same as (b), but for Ti (hole doped), Nb (isovalent), and Cr (electron doped) substituted $\textrm{Cs}\textrm{V}_{3}\textrm{Sb}_{5}$.
Lower panels show the CDW and superconducting phase diagrams against $x$ and temperature.}
 \end{figure}

To discuss the implications of the present results in relation to the substitution-element dependence of the electronic phase diagram, we briefly review previous studies on CDW in pristine, Nb-substituted, and Ti-substituted $\textrm{Cs}\textrm{V}_{3}\textrm{Sb}_{5}$ (Fig. 4) \cite{LiuPRM2023, LiPRB2022, KatoPRL2022}.
As shown in the schematic of normal-state band diagram in Fig. 4(b), pristine $\textrm{Cs}\textrm{V}_{3}\textrm{Sb}_{5}$ has the Sb band at the $\Gamma$ point, and kagome-lattice bands with SP1 and SP2 slightly above and below $E_{\textrm{F}}$, respectively, near the M point.
Based on the largest CDW gap opening ($\sim$70 meV) on the SP1 band below $T_{\textrm{CDW}}$ ($\sim$20 meV on the SP2 band and $\sim$0 meV on the Sb band), electron scattering between SP1 bands at different M points [as connected by cyan arrows in Fig. 4(c)] was suggested to be crucial for the occurrence of CDW \cite{NakayamaPRB2021}.
When V is substituted with Ti, Nb, or Cr [Fig. 4(a)], CDW is monotonically suppressed for all cases \cite{YangSB2022, LiuPRM2023, LiPRB2022, DingPRB2022, PengarXiv2024}.
Regarding the band structure, SP1 shifts away from $E_{\textrm{F}}$ for both Nb and Ti substitutions [Figs. 4(d) and 4(e)].
From this observation, the suppression of CDW was explained by the weakening of inter-SP1 scattering \cite{LiuPRM2023, LiPRB2022, KatoPRL2022}.
However, the present study indicates that considering only SP1 is not sufficient to understand the phase diagram of Cr-substituted $\textrm{Cs}\textrm{V}_{3}\textrm{Sb}_{5}$ because CDW is still suppressed despite SP1 moving closer to $E_{\textrm{F}}$ by electron doping due to Cr substitution [Fig. 4(f)].
To explain these contradictory behaviors, it is suggested that the presence of both SP1 and SP2 in close proximity to $E_{\textrm{F}}$ is the most favorable situation for CDW (it is noted that, similarly to the consideration of only SP1, considering only SP2 does not explain the behavior of Nb and Ti substitution that suppresses CDW while moving SP2 closer to $E_{\textrm{F}}$).
This can be understood qualitatively if the peaks of DOS from SP1 and SP2 are energetically close to each other and a total DOS has a peak in the middle between the two.

Regarding superconductivity, $T_{\textrm{c}}$ increases in Nb-substituted $\textrm{Cs}\textrm{V}_{3}\textrm{Sb}_{5}$ within the CDW phase but decreases in Ti- and Cr-substituted $\textrm{Cs}\textrm{V}_{3}\textrm{Sb}_{5}$ \cite{YangSB2022, LiuPRM2023, LiPRB2022, DingPRB2022, PengarXiv2024}.
Previous ARPES measurements of Nb-substituted $\textrm{Cs}\textrm{V}_{3}\textrm{Sb}_{5}$ have suggested that the DOS near the $\Gamma$ point increases due to a downward energy shift of the Sb band [Fig. 4(e)], which may result in higher $T_{\textrm{c}}$ \cite{KatoPRL2022}.
This explanation, based on the Sb-band DOS, seems to universally explain the trend of $T_{\textrm{c}}$ in the CDW phase of Ti- and Cr-substituted $\textrm{Cs}\textrm{V}_{3}\textrm{Sb}_{5}$.
Specifically, hole doping in the Ti system would lead to a decrease in the DOS around the $\Gamma$ point due to an upward energy shift of the Sb band [Fig. 4(d)], resulting in decrease in $T_{\textrm{c}}$ until the end point of the CDW phase.
In the Cr-substituted case, the orbital-selective electron doping prevents a clear energy shift of the Sb band [Fig. 4(f)].
This peculiar band evolution can explain why the Cr substitution does not result in increase in $T_{\textrm{c}}$.
Nevertheless, there is a point that cannot be explained qualitatively by considering only the Sb band.
That is, in the Ti-substituted case, $T_{\textrm{c}}$ begins to increase once the CDW disappears, whereas $T_{\textrm{c}}$ remains zero in the Cr-substituted case.
This dichotomy may originate from (i) differences in the orbital of the SP band staying near $E_{\textrm{F}}$ ($d_{xz}/d_{yz}$ for SP2 in the Ti-substituted case and $d_{xy}/d_{x^{2}-y^{2}}$ for SP1 in the Cr-substituted case) and/or (ii) stronger pair-breaking effects of the Cr atoms introduced into the kagome lattice compared to the Ti atoms.

In conclusion, we reported a systematic ARPES study of $\textrm{Cs(V}_{1-x}\textrm{Cr}_{x})_{3}\textrm{Sb}_{5}$.
In the normal state above $T_{\textrm{CDW}}$, we found an orbital-selective band shift upon Cr substitution, characterized by the downward energy shift of the V-derived bands in contrast to the stationary nature of the Sb-derived band.
We also found a band doubling due to the 3D CDW only for $x = 0$, whereas the doubling is absent in non-CDW phases of $x = 0.25$ and 0.45.
We have concluded that it is necessary to take into account the presence of two SP bands and their proximity to $E_{\textrm{F}}$ to universally understand the CDW and superconducting phase diagram of chemically substituted $\textrm{Cs}\textrm{V}_{3}\textrm{Sb}_{5}$.
The present results lay a foundation for comprehensively understanding the role of chemical substitution to the phase diagram and relevant electronic states in kagome superconductors.

\begin{acknowledgments}
This work was supported by JST-CREST (No. JPMJCR18T1), JST-PRESTO (No. JPMJPR18L7), Grant-in-Aid for Scientific Research (JSPS KAKENHI Grant Numbers JP21H04435, JP23KJ0099, and JP23K25812), KEK-PF (Proposal numbers: 2021S2-001, 2024S2-001).
The work at Beijing Institute of Technology was supported by the National Key R\&D Program of China (Grant Nos. 2020YFA0308800, 2022YFA1403400), the Natural Science Foundation of China (Grant No. 92065109), and the Beijing Natural Science Foundation (Grant No. Z210006).
S. Suzuki acknowledges support from GP-MS at Tohoku University.
T.K. acknowledges support from GP-Spin at Tohoku University and JST-SPRING (No. JPMJSP2114).
Z.W. thanks the Analysis \& Testing Center at BIT for assistance in facility support.
\end{acknowledgments}

\end{document}